\documentclass[aps,prc,preprint,showpacs]{revtex4}
\usepackage{epsfig}

\def\dfrac#1#2{{\displaystyle\frac{#1}{#2}}}

\begin{document}

%\preprint{}

\title{Two-step contribution to the spin-longitudinal and transverse 
       cross sections of
       the quasielastic ($p$,$n$) reactions }


\author{Yasushi Nakaoka}
%\thanks{}
\affiliation{
        {\it Department of Physics,}
        {\it University of Tokyo,}
        {\it Hongo, Bunkyo-ku,}
        {\it Tokyo 113-0033,} Japan }


\date{Received 31 July 2001; published 11 June 2002}

\begin{abstract}
\quad 
The two-step contribution to the spin-longitudinal and the 
spin-transverse cross sections of $^{12}$C,$^{40}$Ca($p$,$n$) reactions at 
494 MeV and 346 MeV is calculated.
We use a plane-wave approximation and evaluate the relative contributions from 
the one-step and the two-step processes.
We found that the ratios of the two-step to the one-step processes are larger 
in the spin-transverse cross sections than in the spin-longitudinal ones.
Combining these results with the distorted-wave impulse approximation (DWIA) 
results we obtained considerable two-step contributions to the 
spin-longitudinal and the spin-transverse cross sections.
The two-step processes are important in accounting for the underestimation 
of the DWIA results for the spin-longitudinal and the spin-transverse cross 
sections.
\end{abstract}

%\keywords{}

\maketitle

\section{Introduction}
In nucleon induced intermediate energy reactions, a single-step 
reaction model is the simplest way to understand the complex interaction 
between nucleon and nucleus.
However, it is insufficient to explain the cross sections at large 
angles \cite{Luo,Watanabe,Tamura}.
Various groups \cite{Tamura,Ogata} introduced multistep direct processes 
to compensate for the lack of cross sections in these regions.
They calculated angular distributions of the two-step \cite{Tamura} 
and three-step direct processes \cite{Ogata} and found that multistep 
direct processes contribute significantly towards the understanding of 
nucleon induced intermeditate energy reactions.

The two-step processes have also been applied \cite{DePace,Nakaoka} 
to the long-standing $R_{\rm L}/R_{\rm T}$ problem in spin-isospin excitations.
With the random phase approximation (RPA), the spin-longitudinal 
response function $R_{\rm L}$ is enhanced and the peak of its energy spectrum 
is shifted downwards, while the spin-transverse response function $R_{\rm T}$ 
is quenched and its peak is shifted upwards \cite{Alberico}.
The theoretical ratio $R_{\rm L}/R_{\rm T}$ should be greater than 1.
These quantities are extracted experimentally from the measurement of 
the polarization transfer coefficients 
\cite{McClelland,Chen,Taddeucci,Wakasa}.
With these observables, the ratio is found to be less than 1 and is 
inconsistent with theoretical predictions.

\begin{figure}[ht]
   \epsfxsize = 8.0cm
   \centerline{\epsfbox{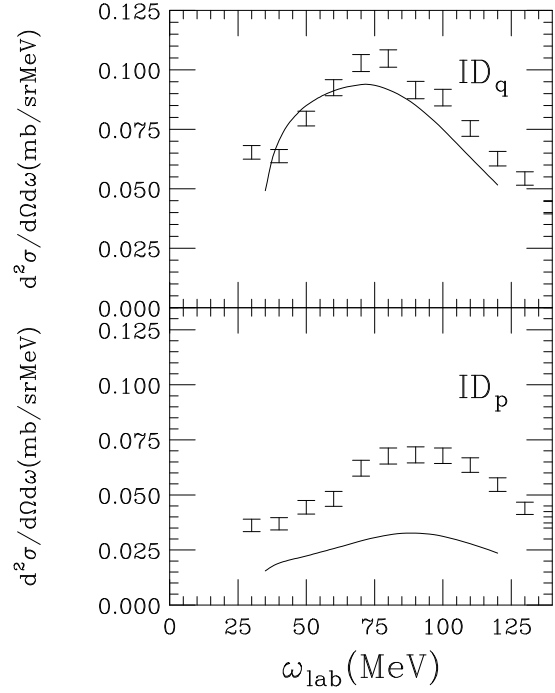}}
   \caption[]
           {The DWIA results for the $^{12}$C($p$,$n$) reaction at 494 MeV
            \cite{Kawahigashi}.
            The upper panel shows the spin-longitudinal cross section, and 
            the lower panel shows the spin-transverse one.
            The horizontal axis is the energy transfer in the laboratory 
            frame.
            \label{dwia}}
\end{figure}

The spin-longitudinal cross section $ID_q$ \cite{Bleszynski}, which 
corresponds to $R_{\rm L}$ \cite{Ichimura}, is roughly reproduced by the 
distorted-wave impulse approximation (DWIA) with the RPA correlation 
in the lower energy transfer region as shown in Fig.~\ref{dwia}, but as 
the energy transfer increases the theoretical results become smaller than the 
experimental results \cite{Kawahigashi}.
In the spin-transverse cross section $ID_p$, which corresponds to $R_{\rm T}$, 
theoretical estimates reproduce only about half of the experimental result in 
the whole energy transfer region \cite{Kawahigashi}.

Nakaoka and Ichimura \cite{Nakaoka} calculated the two-step contribution to 
$ID_q$ and $ID_p$ of the $(p,n)$ reactions within the framework of the 
plane-wave approximation to evaluate the relative contributions from the 
one- and the two-step processes.
Theoretical results including two-step contributions were closer to the 
experimental results than the DWIA predictions for $ID_q$, but they still 
underestimated $ID_p$.

In this formalism we assumed that the path lengths of the incident 
particles in the target nucleus for the one-step and the two-step processes 
are nearly equal, and the effects of absorption in the one-step and the 
two-step processes are similar.
Since we wanted to evaluate the relative contributions from the one-step and 
the two-step processes, we removed the effect of absorption from both 
processes.

Because of the singularity that appears in the Green's function in the 
two-step processes, numerical integration is difficult.
An on-energy shell approximation to the Green's function was used as shown 
in our previous paper \cite{Nakaoka}.
However, since this approximation may not be reliable it is not used herein.

In this paper, we utilize the finite $\epsilon$ in the denominator of the 
Green's function in order to avoid the singularity and calculate the numerical 
integration beyond the on-energy shell in the transferred momentum space.
Then the results are obtained by extrapolation in the limit of zero $\epsilon$.
In Sec.~\ref{Formalism}, we briefly review the formalism of the one-step 
and the two-step cross sections.
In Sec.~\ref{Numerical calculation}, we present the practical method of 
numerical calculation.
In Sec.~\ref{Results}, the results of the one-step and the two-step cross 
sections in the plane-wave calculation are shown.
The two-step cross sections with distortion are roughly estimated using 
the DWIA results and they are compared with experimental results.
In Sec.~\ref{Discussion}, we analyze the numerical results.
Finally, these results are summarized in Sec.~\ref{Summary}.

\section{Formalism}\label{Formalism}
\setcounter{equation}{0}
Consider ($p$,$n$) reactions in which the target is excited to the 
continuum region, for which the multiple scattering formalism of Kerman, 
McManus and Thaler(KMT)~\cite{Kerman} may be used.
In order to see the relative contributions from the one-step and the two-step 
processes, we use a plane-wave approximation.
The total Hamiltonian $H$ is written as
\begin{eqnarray}
H&=&H_0+V, \\
H_0&=&H_{\rm p}+H_{\rm T},
\end{eqnarray}
where $H_{\rm p}$ and $H_{\rm T}$ are the projectile and the target 
Hamiltonians, respectively, 
and $V$ is the sum of the two-body interactions between the projectile 
and the target nucleons.

\subsection{One-step processes}
The one-step $T$ matrix is written in the KMT formalism as
\begin{equation}
T_{n0}^{(1)}({\bf k}_{\rm f},{\bf k}_{\rm i}) \equiv
  \frac{A}{A-1}
  \langle {\bf k}_{\rm f}|\langle \Phi_n|
  \frac{A-1}{A}\sum_{i=1}^A t_i
  |\Phi_0 \rangle|{\bf k}_{\rm i}\rangle ,
\end{equation}
where $|\Phi_n \rangle$ are the eigenstates of the target nucleus with the 
excitation energy $E_n^{\rm int}$; ${\bf k}_{\rm i}$, ${\bf k}_{\rm f}$ 
are the incident and final momenta, respectively; $t_i$ is the 
transition matrix for the projectile and the $i$ th nucleon in the target; 
and $A$ is the target mass number.

In the impulse approximation, $t_i$ is approximated by the free 
nucleon-nucleon transition matrix~(NN $t$ matrix) $t_{\rm NN}$.
Further in the $t\rho$ approximation the one-step $T$ matrix is rewritten as 
\begin{equation}
T_{n0}^{(1)}({\bf k}_{\rm f},{\bf k}_{\rm i})=
  t_{\rm NN}({\bf q}) \langle \Phi_n|\rho({\bf q})|\Phi_0 \rangle,
  \hspace{0.5cm} n\neq0,
\end{equation}
where ${\bf q}={\bf k}_{\rm f}-{\bf k}_{\rm i}$ is the transferred momentum, 
and 
$\rho({\bf q})\equiv \displaystyle \sum_{i=1}^A e^{-i{\bf q}\cdot{\bf r}_i}$ 
is the density operator in the momentum space without the spin dependence.

To analyze the spin observables we inevitably treat the spin dependence 
explicitly.
We introduce the unit vectors
\begin{equation}
\hat{\bf q}=\frac{\bf q}{|\bf q|}, \ \ \
\hat{\bf n}=\frac{{\bf k}_{\rm i}\times{\bf k}_{\rm f}}
                 {|{\bf k}_{\rm i}\times{\bf k}_{\rm f}|}, \ \ \
\hat{\bf p}=\hat{\bf q}\times\hat{\bf n}.
\end{equation}
In the coordinate system $[\hat{\bf q},\hat{\bf n},\hat{\bf p}]$ the 
NN $t$ matrices can be decomposed as
\begin{equation}
t_{\rm NN}({\bf q})=
  \sum_{\mu\bar{\mu}}\sigma_{0\mu}\sigma_{i\bar{\mu}}
  t_{\mu\bar{\mu}}({\bf q}),
\end{equation}
with $\mu,\bar{\mu}=u,q,n,p$, 
where $\sigma_{ju}=I_j,\sigma_{j\mu}=\sigma_j\cdot\hat{\mu}\ (\mu\neq u, 
j=0,i)$.
The operator $\sigma_0$ and $\sigma_i$ denote the spin operators of the 
projectile and the $i$ th nucleon in the target, respectively.

The spin-longitudinal cross section $ID_q$ and the spin-transverse cross 
section $ID_p$ of the ($p$,$n$) reactions can be 
written as \cite{Nakaoka,Bleszynski}
\begin{eqnarray}
ID_q&=& K\frac{\sqrt{s}}{M_{\rm R}}
   t_{q q}^*({\bf q})
   R_{q q}({\bf q},\omega^{\rm int})
   t_{q q}({\bf q}), \\
ID_p&=& K\frac{\sqrt{s}}{M_{\rm R}}
   t_{p p}^*({\bf q})
   R_{p p}({\bf q},\omega^{\rm int})
   t_{p p}({\bf q}), \\
K&\equiv&\frac{\mu_{\rm i}\mu_{\rm f}}{(2\pi)^2}
         \frac{k_{\rm f}}{k_{\rm i}}
\end{eqnarray}
where $M_{\rm R}$ is the invariant mass of the residual nucleus, 
$\sqrt{s}$ is the total energy of the system, 
$\omega^{\rm int}$ is the energy transfer between the intrinsic states, 
and $\mu_{\rm i}\ (\mu_{\rm f})$ is the reduced energy of the projectile
(ejectile).
The factor $|d\omega^{\rm int}/d\omega|=\sqrt{s}/M_{\rm R}$ is the variable 
transformation coefficient between the energy transfer $\omega$ in the center 
of momentum frame and $\omega^{\rm int}$, 
and $R_{\bar{\mu}^\prime \bar{\mu}}({\bf q},\omega^{\rm int})$ is the 
response function of the spin density fluctuation 
\begin{widetext}
\begin{equation}
R_{\bar{\mu}^\prime \bar{\mu}}({\bf q},\omega^{\rm int})\equiv
   \sum_{n\neq0}
   \langle \Phi_0|\rho_{\bar{\mu}^\prime}^\dagger({\bf q})
   |\Phi_n \rangle
   \langle \Phi_n|\rho_{\bar{\mu}}({\bf q})|\Phi_0 \rangle
   \delta(\omega^{\rm int}-E_n^{\rm int}),
\end{equation}
\end{widetext}
where 
$\rho_{\bar{\mu}}({\bf q})\equiv\displaystyle\sum_{i=1}^Ae^{-i{\bf q}\cdot{\bf r}_i} \sigma_{i\bar{\mu}}$.
The response functions $R_{qq}$ and $R_{pp}$ are nothing but the 
spin-longitudinal and the spin-transverse response functions $R_{\rm L}$ and 
$R_{\rm T}$, respectively.
The nuclear correlations are not included in this paper.

\subsection{Two-step processes}
In the two-step processes the $T$ matrix is written as 
\begin{widetext}
\begin{equation}
T_{n0}^{(2)}({\bf k}_{\rm f},{\bf k}_{\rm i})
=\frac{A}{A-1}\sum_{n^\prime\neq0}\int \dfrac{d{\bf k}_{\rm m}}{(2\pi)^3}
\langle{\bf k}_{\rm f}|\langle \Phi_n|\frac{A-1}{A} \sum_{i=1}^A t_i
  |\Phi_{n^\prime} \rangle
|{\bf k}_{\rm m} \rangle
\times
G_{n^\prime}({\bf k}_{\rm m})
\langle{\bf k}_{\rm m}|\langle \Phi_{n^\prime}|
  \frac{A-1}{A}\sum_{i=1}^A t_i
  |\Phi_0\rangle|{\bf k}_{\rm i}\rangle .
\end{equation}
\end{widetext}
Since we adopt a plane-wave approximation, the Green's function is 
diagonal with respect to ${\bf k}_{\rm m}$ as 
\begin{equation}
G_{n^\prime}({\bf k}_{\rm m}) \equiv 
  \dfrac{1}{E -\sqrt{M^2+{\bf k}_{\rm m}^2} 
  -\sqrt{(M_{\rm T}+E_{n^\prime}^{\rm int})^2+{\bf k}_{\rm m}^2}
  +i\epsilon } ,
\end{equation}
where $M$ is the nucleon mass and $M_{\rm T}$ is the target mass.

The target gets excited after the first collision, but assuming that the 
particle-hole pair created in the first collision has nothing to do with 
the second collision, we get $ID_q$ and $ID_p$ as \cite{Nakaoka}
\begin{widetext}
\begin{eqnarray}
ID_q&=&K
    \int d\omega_1
    \int \dfrac{d{\bf q}_1}{(2\pi)^3}
    \int \dfrac{d{\bf q}_1^\prime}{(2\pi)^3}
    \sum_{\mu_2\mu_2^\prime}\sum_{\mu_1\mu_1^\prime}
    X_{\mu_2^\prime \mu_2\mu_1^\prime \mu_1}^{(2)}
    \frac{1}{2}{\rm Tr}
    (\sigma_{0\mu_1^\prime}\sigma_{0\mu_2^\prime}\sigma_q)
    \frac{1}{2}{\rm Tr}
    (\sigma_q\sigma_{0\mu_2}\sigma_{0\mu_1}), \label{eq:idq2} \\
ID_p&=&K
    \int d\omega_1
    \int \dfrac{d{\bf q}_1}{(2\pi)^3}
    \int \dfrac{d{\bf q}_1^\prime}{(2\pi)^3}
    \sum_{\mu_2\mu_2^\prime}\sum_{\mu_1\mu_1^\prime}
    X_{\mu_2^\prime \mu_2\mu_1^\prime \mu_1}^{(2)}
    \frac{1}{2}{\rm Tr}
    (\sigma_{0\mu_1^\prime}\sigma_{0\mu_2^\prime}\sigma_p)
    \frac{1}{2}{\rm Tr}
    (\sigma_p\sigma_{0\mu_2}\sigma_{0\mu_1}), \label{eq:idp2}
\end{eqnarray}
respectively, where ${\bf q}_1={\bf k}_{\rm m}-{\bf k}_{\rm i}$, $\omega_1$ 
is the first-step energy transfer in the center of mass frame, and 
\begin{eqnarray}
X_{\mu_2^\prime \mu_2\mu_1^\prime \mu_1}^{(2)}&\equiv&
  \frac{s}{M_{\rm R}^2}
  \left( \frac{A-1}{A} \right)^2
  \sum_{\bar{\mu}_2\bar{\mu}_2^\prime}
  \sum_{\bar{\mu}_1\bar{\mu}_1^\prime}
  t_{\mu_2^\prime\bar{\mu}_2^\prime}^*({\bf q}-{\bf q}_1^\prime)
  t_{\mu_2\bar{\mu}_2}({\bf q}-{\bf q}_1)
  R_{\bar{\mu}_2^\prime \bar{\mu}_2}
    ({\bf q}-{\bf q}_1,{\bf q}-{\bf q}_1^\prime;
    \omega^{\rm int}-\omega_1^{\rm int}) \nonumber \\
&\times&
  G^{*}({\bf k}_{\rm i}+{\bf q}_1^\prime;\omega_1^{\rm int})
  G({\bf k}_{\rm i}+{\bf q}_1;\omega_1^{\rm int})
  t_{\mu_1^\prime\bar{\mu}_1^\prime}^*({\bf q}_1^\prime)
  t_{\mu_1\bar{\mu}_1}({\bf q}_1)
  R_{\bar{\mu}_1^\prime \bar{\mu}_1}
    ({\bf q}_1,{\bf q}_1^\prime;\omega_1^{\rm int}) \label{eq:X2}
\end{eqnarray}
\end{widetext}
with $\mu_1,\bar{\mu}_1=u,q_1,n_1,p_1$ and $\mu_2,\bar{\mu}_2=u,q_2,n_2,p_2$.
Here we define the unit vectors 
\begin{equation}
\hat{\bf q}_2=\frac{{\bf q}-{\bf q}_1}{|{\bf q}-{\bf q}_1|}, \hspace{0.3cm}
\hat{\bf n}_2=\frac{{\bf k}_{\rm f}\times\hat{\bf q}_2}
                  {|{\bf k}_{\rm f}\times\hat{\bf q}_2|}, \hspace{0.3cm}
\hat{\bf p}_2=\hat{\bf q}_2\times\hat{\bf n}_2.
\end{equation}
The excitation energy in the Green's function was replaced with the 
first-step intrinsic energy transfer $\omega_1^{\rm int}$, and the response 
function is extended to a nondiagonal form with respect to the transferred 
momentum ${\bf q}$ as
\begin{widetext}
\begin{equation}
R_{\bar{\mu}^\prime \bar{\mu}}({\bf q},{\bf q}^\prime;\omega^{\rm int})=
\sum_{n\neq0}
\langle \Phi_0|\rho_{\bar{\mu}^\prime}^\dagger({\bf q}^\prime)|\Phi_n\rangle
\langle \Phi_n|\rho_{\bar{\mu}}({\bf q})|\Phi_0\rangle
\delta(\omega^{\rm int}-E_n^{\rm int}) .
\label{eq:response}
\end{equation}
\end{widetext}
The factor $(A-1)/A$ in Eq.~(\ref{eq:X2}) represents that the struck nucleon 
in the first step is never struck again in the second step.
 
The interference terms with $T_{n0}^{(1)}$ and $T_{n0}^{(2)}$ are neglected. 
Exact target states are the sum of 1$p$-1$h$ states, 2$p$-2$h$ states, and 
so on.
The coefficients of 1$p$-1$h$ states are expected to be random with respect to 
the exact target states for a fixed 1$p$-1$h$ state.
Then the target states can be represented by 1$p$-1$h$ states with a 
statistical distribution function.

It can be considered that the different order $T$ matrices mainly excite 
different number particle-hole states. 
Then, the final state of $T_{n0}^{(1)}$ is different from that of 
$T_{n0}^{(2)}$. 
Hence the contribution from the interference terms can be neglected.

\begin{figure}[ht]
   \epsfxsize = 10.0cm
   \centerline{\epsfbox{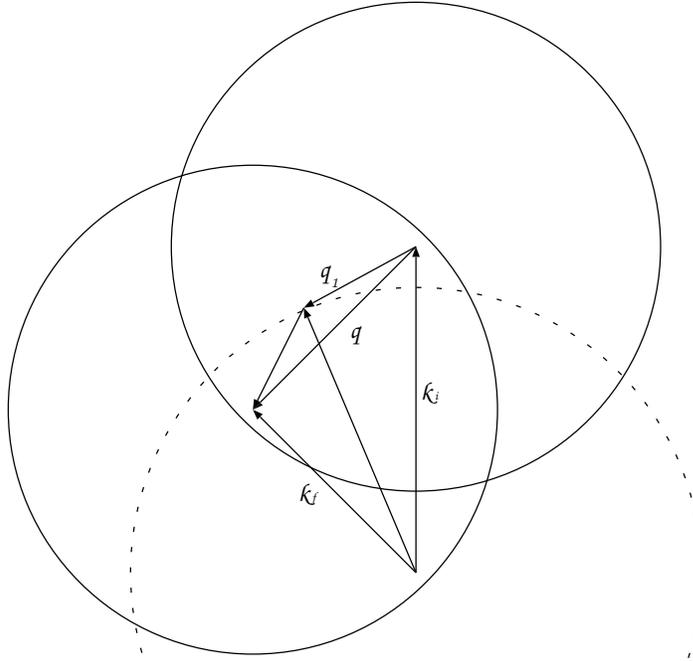}}
   \caption[]
           {The integration region in the transferred momentum space.
            Two spheres drawn with solid lines have a radius 4.8(1/fm) 
            for $^{12}$C,
            beyond which the response functions are negligible.
            The dashed sphere is the on-energy shell.
            \label{rgn}}
\end{figure}

In Fig.~\ref{rgn} the integration region in the transferred momentum 
space is shown.
Outside the spheres drawn with solid lines the values of the response 
functions are negligible.
With this property of the response functions the integration region is 
restricted within the overlap region of the two spheres.

In the limit of $\epsilon=0$, the Green's function is rewritten as
\begin{eqnarray}
&G&({\bf k}_{\rm m};\omega_1^{\rm int})={\cal P}
   \frac{1}{E-\sqrt{M^2+{\bf k}_{\rm m}^2}
            -\sqrt{(M_{\rm T}+\omega_1^{\rm int})^2+{\bf k}_{\rm m}^2}} 
  \nonumber \\
&-&i\pi\delta\left(E-\sqrt{M^2+{\bf k}_{\rm m}^2}
             -\sqrt{(M_{\rm T}+\omega_1^{\rm int})^2+{\bf k}_{\rm m}^2}
   \right). 
\end{eqnarray}
In our previous paper the principal part of the Green's function is neglected
(the on-energy shell approximation).
With the $\delta$ functions the integration region was restricted onto the 
surface of the sphere drawn with a dashed line in Fig.~\ref{rgn}.

Since the validity of this approximation is questionable it is removed.
We first calculate the sevenfold integration in Eqs.~(\ref{eq:idq2}) and 
(\ref{eq:idp2}) for several finite values of $\epsilon$ in the Green's 
function to avoid the singularity that appears at $\epsilon=0$.
We then extrapolate the results to those at $\epsilon=0$.

\section{Numerical calculation}\label{Numerical calculation}
\subsection{Suppression of energy transfer dependence}

In Ref.~\cite{Nakaoka} the process in which the incident particle loses the 
energy $\omega_1$ in the first collision and propagates in the intermediate 
state at the energy that is smaller by $\omega_1$ than the incident energy 
is described, but in this paper the $\omega_1$ dependence is removed 
from the NN $t$ matrices in the second step and from the Green's functions to 
make the sevenfold integration easier.

\begin{figure}[ht]
   \epsfxsize =  8.0cm
   \centerline{\epsfbox{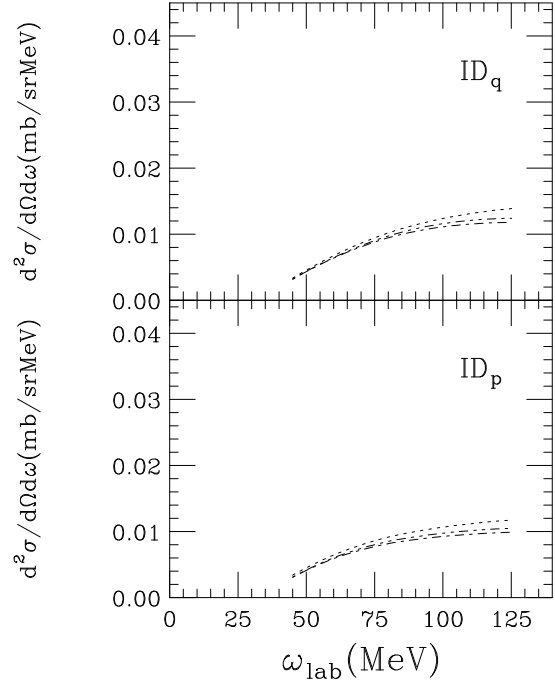}}
   \caption[No $\omega_1$ dependence]
           {The dependence of the first-step energy transfer $\omega_1$ 
            for $^{12}$C($p$,$n$) reaction at 494 MeV.
            The dotted lines indicate the two-step cross sections with 
            the full $\omega_1$ dependence. 
            The dotted-dashed lines are those without the $\omega_1$ 
            dependence in NN $t$ matrices, and the dotted-dotted-dashed 
            lines represent the two-step cross sections without the 
            $\omega_1$ dependence both in the NN $t$ matrices and the 
            Green's function, respectively.
            \label{cs2now}}
\end{figure}

To justify this approximation we show in Fig.~\ref{cs2now} the $\omega_1$ 
dependence of the two-step cross sections with the on-energy shell 
approximation.
The removal of the $\omega_1$ dependence from the NN $t$ matrices slightly 
decreases the two-step cross sections for both $ID_q$ and $ID_p$.
Further removal of the $\omega_1$ dependence from the Green's functions 
makes them increase a little.
In spite of these procedures the total variations in the two-step cross 
sections with the on-energy shell approximation are negligible.

The suppression is also expected to have little effect on the estimation of 
the two-step cross sections without the on-energy shell approximation, 
because the on-energy shell approximation is a restriction of the integration 
region with respect to the transferred momentum space and has nothing to do 
with the integration of the energy transfer $\omega_1$.

\subsection{Parameters and mesh sizes}
The two-step cross sections are calculated at 
$\epsilon=5.0$, 10.0, 15.0, and 20.0 MeV.
The mesh sizes of the numerical integration are almost the same as those in 
Ref.~\cite{Nakaoka}, although the $z$ axis is set parallel to the 
direction of ${\bf q}$.
$\Delta \omega_1^{\rm int}=10.0$ MeV, $\Delta q_1^{\rm int}=0.2$(1/fm), 
where $q_1^{\rm int}=\{(A-1)/A\}q_1$, and $\Delta \cos \theta_1=0.125$.
We set $\Delta \phi_1=\pi/96,~\pi/48$ for $\epsilon=5.0$, and 10.0 MeV, 
respectively, and $\Delta \phi_1=\pi/24$ for $\epsilon=15.0$, and 20.0 MeV.

\section{Results}\label{Results}
Figure \ref{cs2W} shows the two-step cross sections of $^{12}$C($p$,$n$) 
reaction at 494 MeV as a function of $\epsilon$ at the energy transfer of 
85 MeV.
By extrapolating the calculated two-step cross sections toward 
$\epsilon=0.0$ MeV, they are expected to be 1.90 and 1.93 times as much as 
the results with the on-energy shell approximation.

\begin{figure}[ht]
   \epsfxsize = 8.0cm
   \centerline{\epsfbox{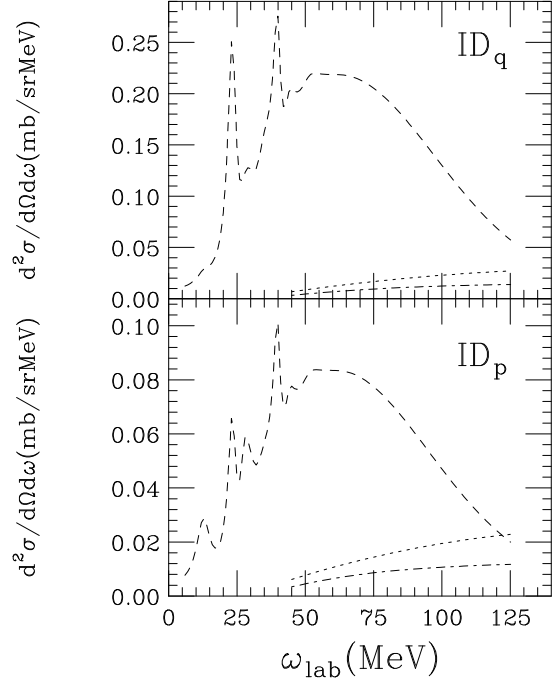}}
   \caption[]
           {The cross sections of $^{12}$C($p$,$n$) reaction at 494 MeV as 
            a function of $\omega_{\rm lab}$.
            The two-step cross sections are indicated with dotted lines, 
            and they are extrapolated to $\epsilon=0.0$ MeV. 
            The dotted-dashed lines are those with the on-energy shell
            approximation.
            The dashed lines are the one-step cross sections.
            \label{csc12k494a18pn}}
\end{figure}

Figure \ref{csc12k494a18pn} shows the cross sections of $^{12}$C($p$,$n$) 
reaction at 494 MeV as a function of the energy transfer in the laboratory 
frame, $\omega_{\rm lab}$, where the two-step cross sections without the 
on-energy shell approximation are extrapolated to $\epsilon=0.0$ MeV.
The two-step cross sections increase as the energy transfer increases.
The results with the on-energy shell approximation are shown with 
dotted-dashed lines.

The ratios of the two-step cross sections to the one-step ones at 
$\omega_{\rm lab}=125$ MeV are about $47\%$ and $114\%$ for $ID_q$ and $ID_p$, 
respectively.
Those with the on-energy shell approximation are revised to include the factor 
$(A-1)^2/A^2$ in Eq.~(\ref{eq:X2}) and they are $24\%$ and $59\%$, 
respectively.
The relative contribution is much larger in $ID_p$ than in $ID_q$.
These data are displayed in Table \ref{tab:ratio}.

\begin{figure}[ht]
   \epsfxsize = 8.0cm
   \centerline{\epsfbox{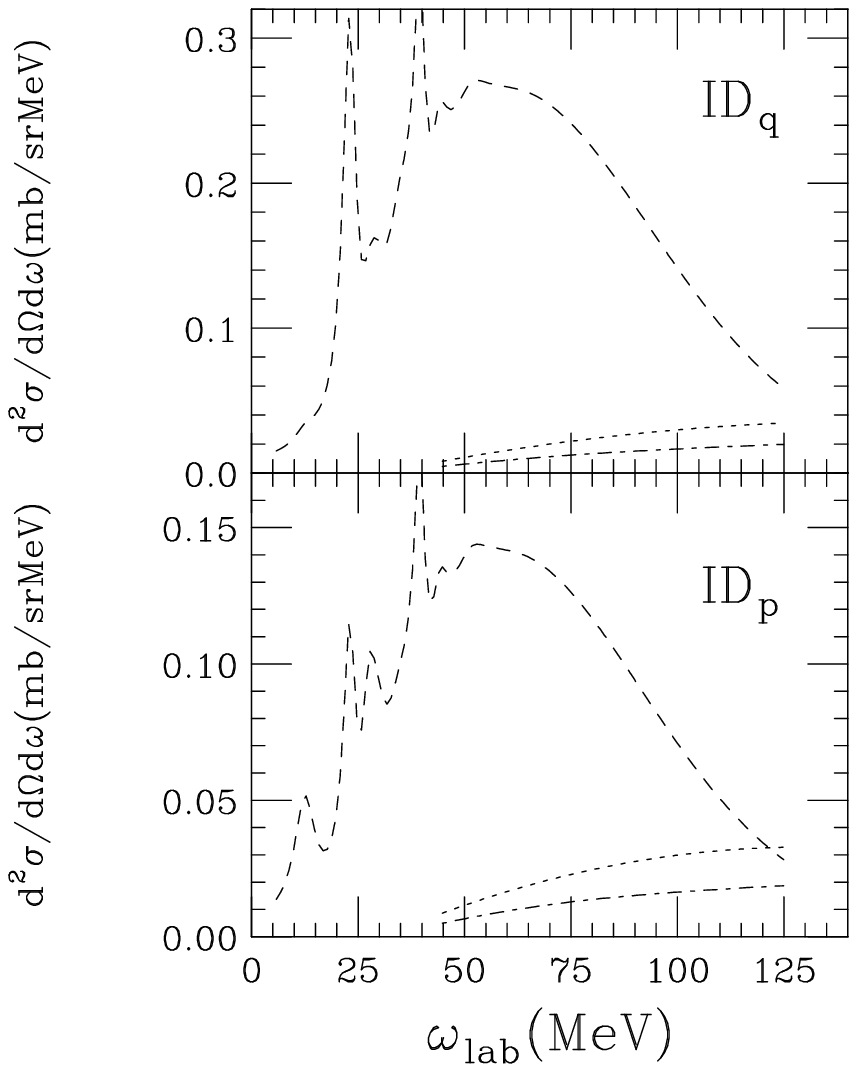}}
   \caption[]
           {Same as Fig.~\ref{csc12k494a18pn}, but at 346 MeV.
            \label{csc12k346a22pn}}
\end{figure}

The cross sections of $^{12}$C($p$,$n$) reaction at 346 MeV are shown in 
Fig.~\ref{csc12k346a22pn}.
The ratios are $59\%$ and $116\%$ at $\omega_{\rm lab}=125$ MeV, and those 
with the on-energy shell approximation are $34\%$ and $66\%$.

\begin{figure}[ht]
   \epsfxsize = 8.0cm
   \centerline{\epsfbox{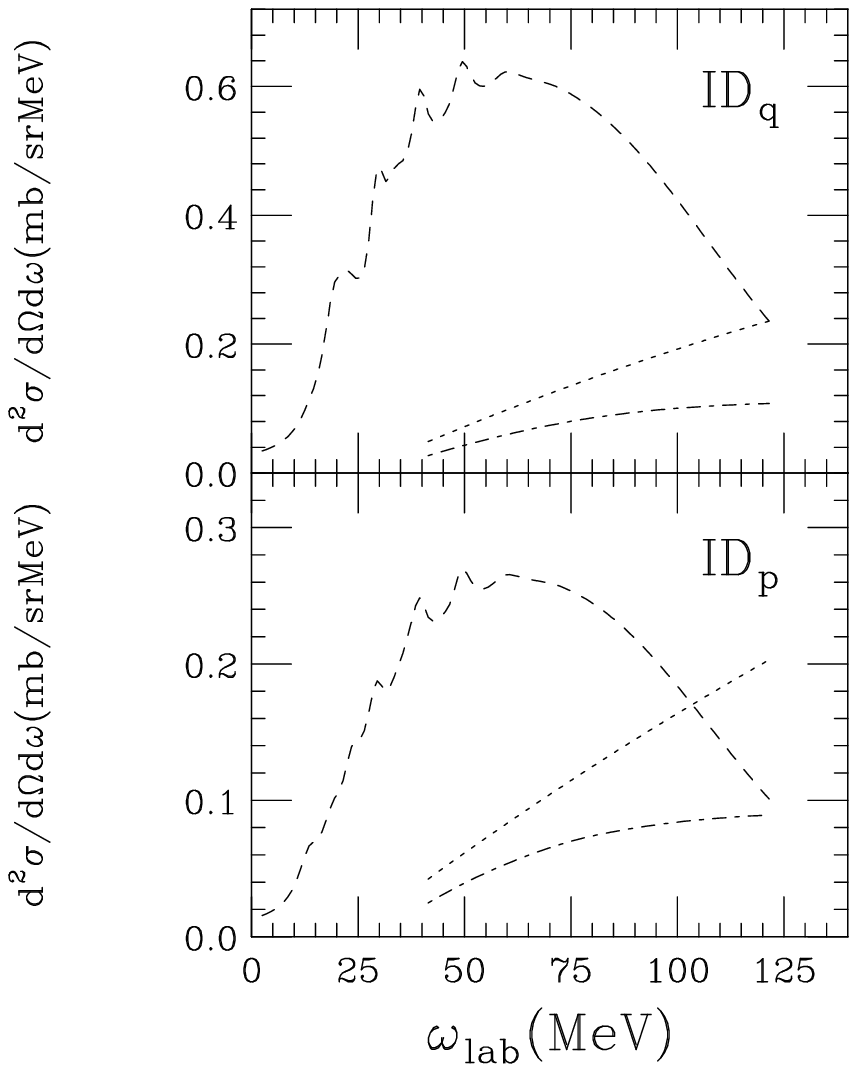}}
   \caption[]
           {Same as Fig.~\ref{csc12k494a18pn}, but for $^{40}$Ca.}
   \label{csca40k494a18pn}
\end{figure}

\begin{figure}[ht]
   \epsfxsize = 8.0cm
   \centerline{\epsfbox{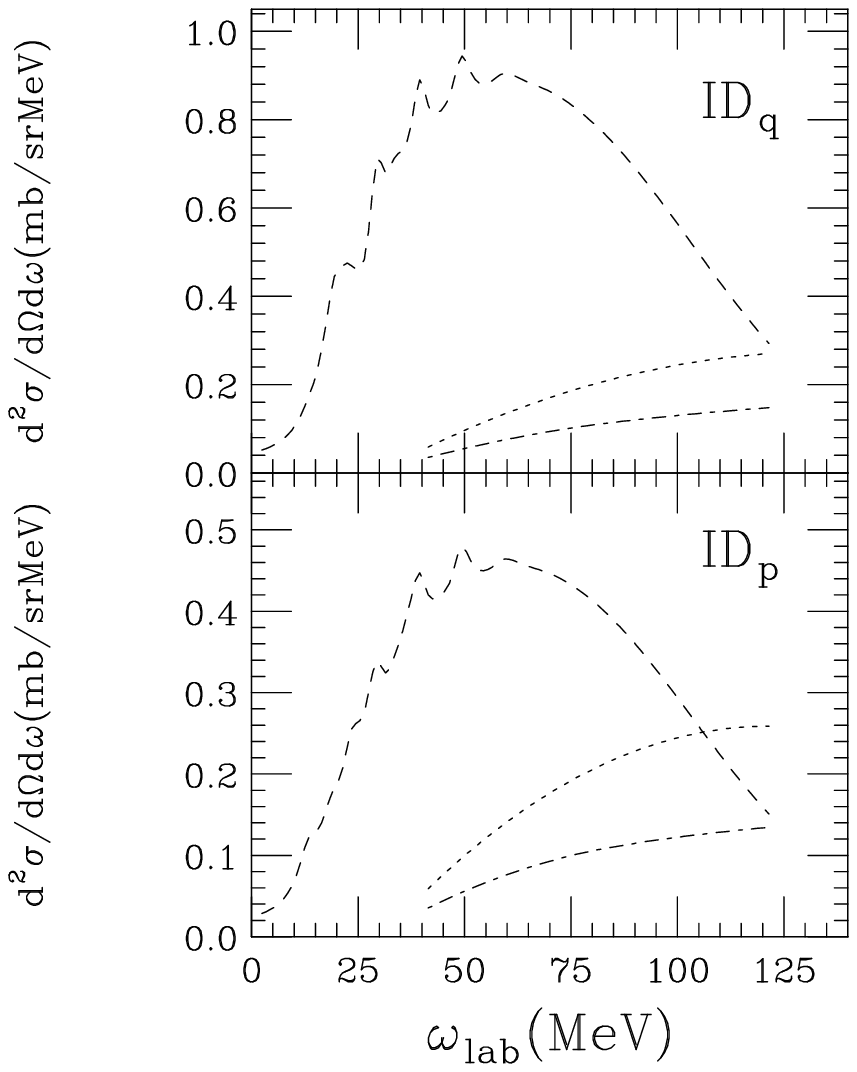}}
   \caption[]
           {Same as Fig.~\ref{csca40k494a18pn}, but at 346 MeV.
            \label{csca40k346a22pn}}
\end{figure}

The cross sections of $^{40}$Ca($p$,$n$) reactions at 494 MeV and at 346 MeV 
are shown in Figs.~\ref{csca40k494a18pn} and \ref{csca40k346a22pn}, 
respectively.
The ratios at $\omega_{\rm lab}=121$ MeV are shown in Table \ref{tab:ratio}.

Since the two-step cross sections are calculated in the plane-wave 
approximation, it is difficult to know whether these results account for the 
underestimation of $ID_q$ and $ID_p$.
One possible way of showing the two-step effects may be to multiply 
the uncorrelated DWIA results \cite{Kawahigashi} with the ratios of the 
two-step cross sections to the one-step cross sections in the plane-wave 
approximation.
Then these renormalized contributions are added to the DWIA results with the 
RPA correlation, 
since the response functions without the RPA correlation were used in the 
plane-wave calculation of the one-step and the two-step cross sections.

\begin{figure}[ht]
   \epsfxsize = 8.0cm
   \centerline{\epsfbox{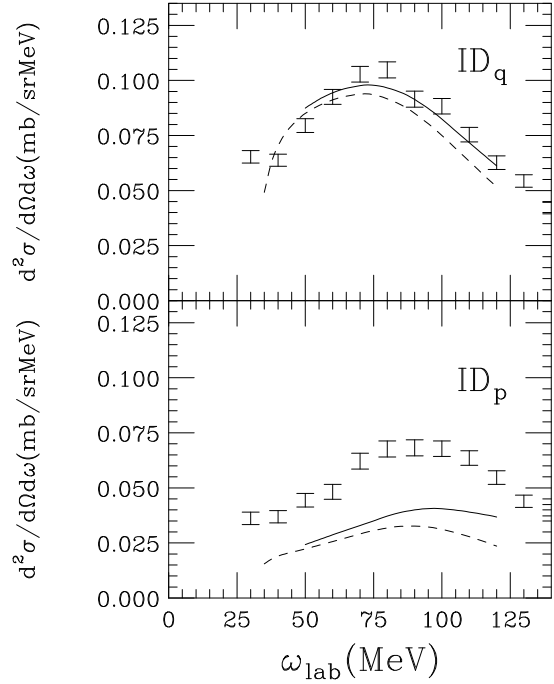}}
   \caption[]
           {The DWIA \cite{Kawahigashi} with the two-step results 
            for the $^{12}$C($p$,$n$) reaction at 494 MeV.
            The upper panel shows $ID_q$, and the lower panel does $ID_p$.
            The dashed lines indicate the DWIA results.
            The solid lines include the two-step effects.
            Experimental data are taken from Ref.~\cite{Taddeucci}.
            \label{dw0c12k494a18pn}}
\end{figure}

The two-step contributions for the $^{12}$C($p$,$n$) reaction at 494 MeV thus 
obtained are shown in Fig.~\ref{dw0c12k494a18pn}.
The numerical results get closer to the experimental result both in $ID_q$ 
and $ID_p$ than the previous estimation with the on-energy shell 
approximation.
The two-step cross sections have contributions comparable to the discrepancy 
between the DWIA results and the experimental ones.

\begin{figure}[hbt]
   \epsfxsize = 8.0cm
   \centerline{\epsfbox{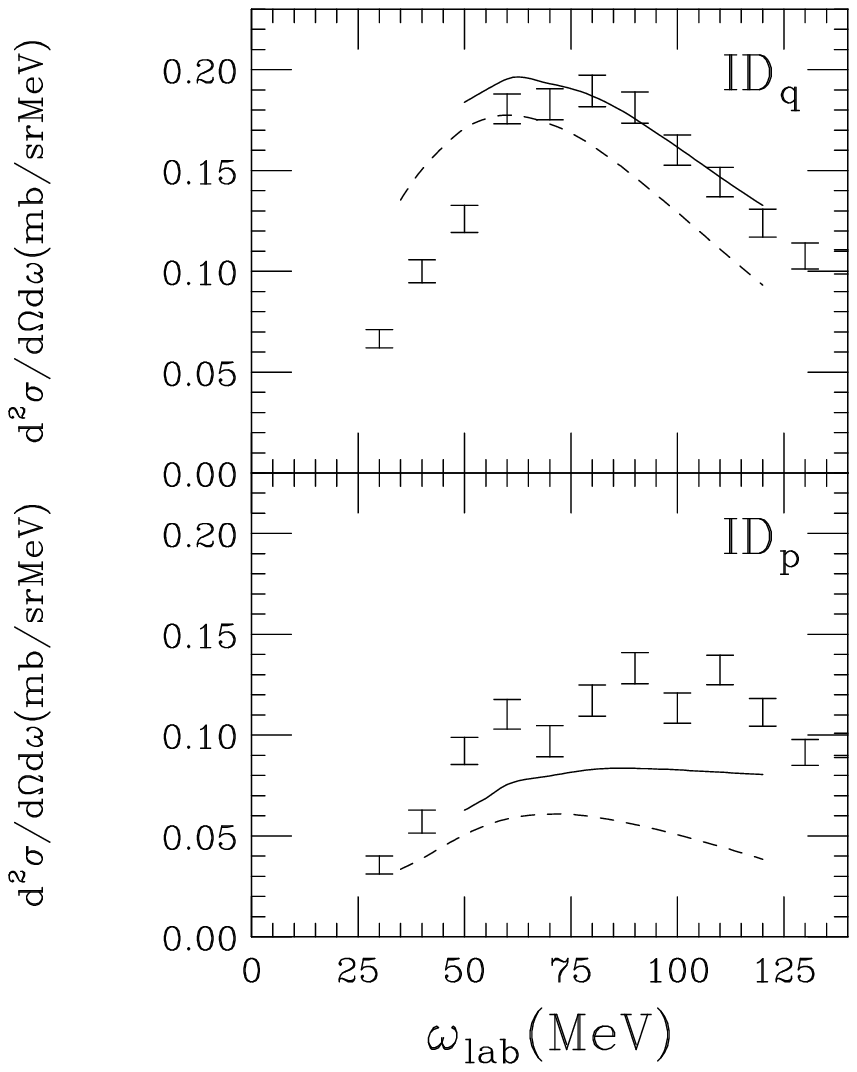}}
   \caption[]
           {Same as Fig.~\ref{dw0c12k494a18pn}, but for $^{40}$Ca.
            \label{dw0ca40k494a18pn}}
\end{figure}

The result for the $^{40}$Ca($p$,$n$) reaction at 494 MeV is displayed in 
Fig.~\ref{dw0ca40k494a18pn}.
The calculated results approach the experimental results both in $ID_q$ and 
$ID_p$ and account for a significant portion of the previously underestimated 
spin-transverse cross section.
Here one must note that the present estimation of the two-step contribution 
with distortion is crude, but Figs.~\ref{dw0c12k494a18pn} and 
\ref{dw0ca40k494a18pn} indicate to what extent the two-step processes are 
important to account for the discrepancy between the DWIA results and the 
experimental results.

\section{Discussion}\label{Discussion}
In this section a qualitative explanation for the origin of the 
difference between the results with and without the on-energy shell 
approximation is considered by first defining 
\begin{widetext}
\begin{eqnarray}
\tilde{X}^{(2)}(\omega_1^{\rm int})&\equiv&
  \int \dfrac{d{\bf q}_1}{(2\pi)^3}
  \int \dfrac{d{\bf q}_1^\prime}{(2\pi)^3}
  t_{\rm NN}({\bf q}-{\bf q}_1)
  t_{\rm NN}^*({\bf q}-{\bf q}_1^\prime)
  R({\bf q}-{\bf q}_1,{\bf q}-{\bf q}_1^\prime;
    \omega^{\rm int}-\omega_1^{\rm int}) \nonumber \\
&\times&
  G({\bf k}_{\rm i}+{\bf q}_1)
  G^{*}({\bf k}_{\rm i}+{\bf q}_1^\prime)
  t_{\rm NN}({\bf q}_1)t_{\rm NN}^*({\bf q}_1^\prime)
  R({\bf q}_1,{\bf q}_1^\prime;\omega_1^{\rm int}), \label{eq:X2w}
\end{eqnarray}
\end{widetext}
which is Eq.~(\ref{eq:idq2}) or (\ref{eq:idp2}) but the integration 
with $\omega_1$ is not included.
Here we suppressed the spin dependence for simplicity.
Because the on-energy shell approximation is a restriction of the 
integration region in the transferred momentum space, the accuracy of the 
on-energy shell approximation is considered only with regards to the 
integrations of the transferred momenta.

The quantity $\tilde{X}^{(2)}$ has to be the square of the absolute 
value of a complex number, because the cross section is obtained from 
the square of the $T$ matrix.
Actually the response function is the sum of target excited states as 
written in Eq.~(\ref{eq:response}) and each term is divided into a function 
of ${\bf q}_1$ and a function of ${\bf q}_1^\prime$, 
and Eq.~(\ref{eq:X2w}) is rewritten as the product of the integration of 
${\bf q}_1$ and its complex conjugate, i.e., the integration of 
${\bf q}_1^\prime$.

\begin{figure}[ht]
   \epsfxsize = 8.0cm
   \centerline{\epsfbox{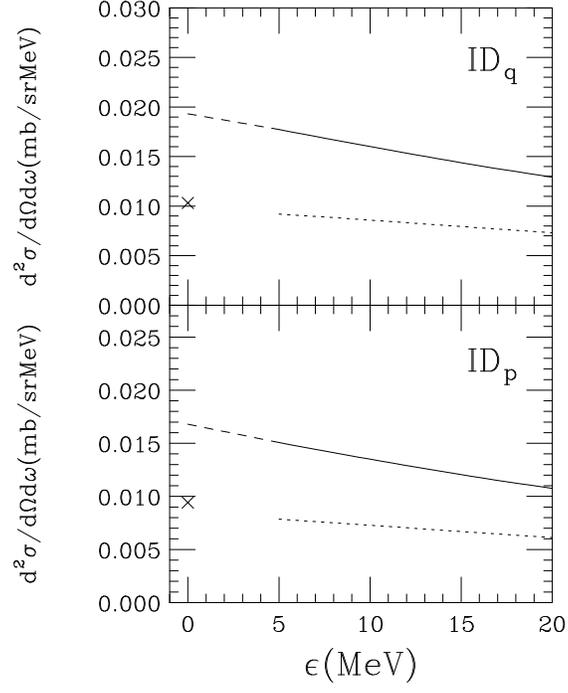}}
   \caption[]
           {The two-step cross sections of $^{12}$C($p$,$n$) reaction 
            at 494 MeV as a function of $\epsilon$.
            The energy transfer is 85 MeV.
            The dashed lines indicate the results extrapolated smoothly
            from the calculated results. 
            The dotted lines do the contributions from the fourth term 
            in Eq.~(\ref{eq:X2w4trm}).
            The results with the on-energy shell approximation are 
            plotted with cross symbols at $\epsilon=0.0$ MeV.
            \label{cs2W}}
\end{figure}

Dividing the Green's functions into the real part and the imaginary part, 
one gets
\begin{widetext}
\begin{eqnarray}
\tilde{X}^{(2)}(\omega_1^{\rm int})&=&
  \int \dfrac{d{\bf q}_1}{(2\pi)^3}
  \int \dfrac{d{\bf q}_1^\prime}{(2\pi)^3}
  t_{\rm NN}({\bf q}-{\bf q}_1)
  t_{\rm NN}^*({\bf q}-{\bf q}_1^\prime)
  R({\bf q}-{\bf q}_1,{\bf q}-{\bf q}_1^\prime;
    \omega^{\rm int}-\omega_1^{\rm int}) \nonumber \\
&\times&
  \{ {\rm Re}G({\bf k}_{\rm i}+{\bf q}_1)
    {\rm Re}G({\bf k}_{\rm i}+{\bf q}_1^\prime)
+i{\rm Im}G({\bf k}_{\rm i}+{\bf q}_1)
    {\rm Re}G({\bf k}_{\rm i}+{\bf q}_1^\prime) \nonumber \\
&-&i{\rm Re}G({\bf k}_{\rm i}+{\bf q}_1)
    {\rm Im}G({\bf k}_{\rm i}+{\bf q}_1^\prime)
+ {\rm Im}G({\bf k}_{\rm i}+{\bf q}_1)
    {\rm Im}G({\bf k}_{\rm i}+{\bf q}_1^\prime) \} \nonumber \\
&\times&
  t_{\rm NN}({\bf q}_1)t_{\rm NN}^*({\bf q}_1^\prime)
  R({\bf q}_1,{\bf q}_1^\prime;\omega_1^{\rm int})
 \label{eq:X2w4trm}
\end{eqnarray}
\end{widetext}
It is the fourth term that remains when the on-energy shell 
approximation is applied.
The same argument as in the previous paragraph can be made with respect to 
the first term and the fourth term, and both of these two terms become 
positive.
The second and the third terms are the complex conjugate of each other, 
and the sum gives a real number.
However, the contributions to $ID_q$ and $ID_p$ from the second and the third 
terms are found to be small.
In Fig.~\ref{cs2W}, they are only $-1.7\times10^{-4}$ and $-2.7\times10^{-4}$ 
out of $1.29\times10^{-2}$ mb and $1.08\times10^{-2}$ mb at $\epsilon=20.0$ 
MeV, respectively.
Therefore the cross sections obtained in Sec.~\ref{Results} are larger than 
those with the on-energy shell approximation.

The contributions from the fourth term are displayed in Fig.~\ref{cs2W} 
with dotted lines.
Their extrapolation toward $\epsilon=0.0$ MeV seem to hit the results with the 
on-energy shell approximation.
This shows that the results obtained in Sec.~\ref{Results} are consistent 
with the previous estimations with the on-energy shell approximation.

\section{Summary}\label{Summary}
Two-step contribution to the intermediate energy ($p$,$n$) reactions were 
calculated using the plane-wave approximation.
We carried out the integration beyond the on-energy shell.
The ratios of two-step to one-step contributions were found to be larger 
for the spin-transverse cross sections than for the spin-longitudinal cross 
sections.
The two-step cross sections without the on-energy shell approximation 
are about twice as large as our previous results with the on-energy shell 
approximation.

The cross sections including two-step processes were compared with 
experimental results.
We conclude that the two-step cross sections have contributions comparable 
to the discrepancy between the DWIA and the experimental results in the highly 
excited region for $ID_q$ and they account for a portion of the previously 
underestimated $ID_p$.

\begin{table}
\caption{The ratios of the two-step cross sections to the one-step ones 
         at 125 MeV for $^{12}$C and at 121 MeV for $^{40}$Ca.
         Values in parentheses are those with the on-energy shell 
         approximation.\label{tab:ratio}}
\begin{tabular}{cccc}
Target      \ $K_{\rm lab}$(MeV) \ $ID_q$($\%$) \ $ID_p$($\%$) \\ \hline
{$^{12}$C}\hspace{7mm}  \ 494\hspace{10mm}\  47(24) \ 114(59) \\
\hspace{12mm}           \ 346\hspace{10mm}\  59(34) \ 116(66) \\
{$^{40}$Ca}\hspace{5mm} \ 494\hspace{8mm} \  96(46) \ 192(88) \\
\hspace{12mm}           \ 346\hspace{8mm} \  92(50) \ 172(89) \\
\end{tabular}
\end{table}

% Surround table environment with turnpage environment for landscape
% table
% \begin{table}
% \begin{ruledtabular}
% \end{tabular}
% \end{ruledtabular}
% \end{table}

%\appendix
%\section{}

%\begin{acknowledgments}
%\end{acknowledgments}
\begin{acknowledgments}
The author would be grateful to Professor Hideyuki Sakai, the University 
of Tokyo, Professor Munetake Ichimura, Hosei University, for useful 
advices, and to Professor Mark B.~Greenfield, International Christian 
University, for improving the English of the manuscript.
\end{acknowledgments}


\begin{thebibliography}{}
\bibitem{Luo}Y.~L.~Luo and M.~Kawai, Phys.~Rev.~C {\bf 43},~2367 (1991).
\bibitem{Watanabe}Y.~Watanabe and M.~Kawai, Nucl.~Phys.~{\bf A560},~43 (1993).
\bibitem{Tamura}T.~Tamura, T.~Udagawa, and H.~Lenske, 
Phys.~Rev.~C {\bf 26},~379 (1982).
\bibitem{Ogata}K.~Ogata, M.~Kawai, Y.~Watanabe, S.~Weili, and M.~Kohno, 
Phys.~Rev.~C {\bf 60},~054605 (1999).
\bibitem{DePace}A.~De Pace, Phys.~Rev.~Lett.~{\bf 75},~29 (1995).
\bibitem{Nakaoka}Y.~Nakaoka and M.~Ichimura, Prog.~Theor.~Phys.~{\bf 102},~599 
(1999).
\bibitem{Alberico}W.~M.~Alberico, M.~Ericson, and A.~Molinari, Nucl.~Phys.~
{\bf A379},~429 (1982).
\bibitem{McClelland}J.~B.~McClelland {\it et al.}, Phys.~Rev.~Lett.~
{\bf 69},~582 (1992).
\bibitem{Chen}X.~Y.~Chen {\it et al.}, Phys.~Rev.~C {\bf 47},~2159 (1993).
\bibitem{Taddeucci}T.~N.~Taddeucci {\it et al.}, 
Phys.~Rev.~Lett.~{\bf 73},~3516 (1994).
\bibitem{Wakasa}T.~Wakasa {\it et al.}, Phys.~Rev.~C {\bf 59},~3177 (1999).
\bibitem{Bleszynski}E.~Bleszynski, M.~Bleszynski, and C.~A.~Whitten, 
Jr., Phys.~Rev.~C {\bf 26},~2063 (1982).
\bibitem{Ichimura}M.~Ichimura and K.~Kawahigashi, 
Phys.~Rev.~C {\bf 45},~1822 (1992).
\bibitem{Kawahigashi}K.~Kawahigashi, K.~Nishida, A.~Itabashi, and M.~Ichimura,
Phys.~Rev.~C {\bf 63},~044609 (2001).
\bibitem{Kerman}
A.~K.~Kerman, H.~McManus, and R.~M.~Thaler, Ann.~Phys.(Leipzig)~{\bf 8} 551 
(1959).
\end{thebibliography}

% \begin{figure}
% \end{figure}


\end{document}